\documentclass[prl,aps,twocolumn,superscriptaddress,floatfix,noshowpacs,nofootinbib,10pt]{revtex4-2}

\usepackage[utf8]{inputenc}     	
\usepackage{amsmath, amssymb, mathtools, mathrsfs}
\usepackage[colorlinks=true]{hyperref}
\hypersetup{colorlinks,linkcolor={red},citecolor={blue},urlcolor={blue}}  
\usepackage{color}
\usepackage{graphicx}
\usepackage[normalem]{ulem}
\usepackage{mathtools}
\usepackage{stmaryrd}
\usepackage{float}
\usepackage{textalpha}

\newcommand{\bs}[1]{{\boldsymbol{#1}}}

\newcommand{\br}{\bs{r}}

\newcommand{\bq}{\bs{q}}
\newcommand{\bp}{\bs{p}}

\begin{document}
\title{Anomalous Landau damping and algebraic thermalization  in  two-dimensional superfluids far from equilibrium}
\author{Cl\'ement Duval}
\email{clement.duval2@cea.fr}
\affiliation{Laboratoire Kastler Brossel, Sorbonne Universit\'{e}, CNRS, ENS-PSL Research University, 
Coll\`{e}ge de France; 4 Place Jussieu, 75005 Paris, France}

\author{Nicolas Cherroret}
\email{nicolas.cherroret@lkb.upmc.fr}
\affiliation{Laboratoire Kastler Brossel, Sorbonne Universit\'{e}, CNRS, ENS-PSL Research University, 
Coll\`{e}ge de France; 4 Place Jussieu, 75005 Paris, France}

\begin{abstract}
We present a quantitative description of the thermalization dynamics of  far-from-equilibrium, two-dimensional (2D) Bose superfluids. Our analysis leverages a quantum kinetic formalism and allows us to identify two successive regimes of relaxation: an initial damping of quasi-particles due to Landau scattering processes, followed by the slower establishment of a global equilibrium at long time. For a far-from-equilibrium initial state, we find that Landau damping differs from the conventional picture of exponentially relaxing quasi-particles. Moreover, our results showcase a pronounced mechanism of algebraic transport at late times, rooted in energy conservation and compatible with 2D diffusion. 
Using theoretical and numerical arguments, we construct a detailed dynamical portrait of global equilibration in 2D superfluids.
\end{abstract}
\maketitle

Understanding the thermalization dynamics of many-body systems near integrability is a key challenge of non-equilibrium quantum physics~\cite{Polkovnikov2011, Eisert2015, Langen2015}. 
Unlike integrable systems, which do not reach thermal equilibrium in the traditional sense due to the existence of an infinite number of integrals of motion \cite{Kinoshita2006, Rigol2007, Calabrese2016, Bouchoule2023}, those close to integrability generically thermalize but often through a non-trivial process. Indeed, their short-time dynamics is first governed by the integrable part of the Hamiltonian, which promotes the system to a long-lived pre-thermal state \cite{Kollar2011, Gring2012, Bertini2015, Mori2018, Mallayya2019}. During this regime, the dynamics is driven by coherent quasi-particles that may give rise to quantum interferences, as observed in recent experiments with quantum fluids \cite{Hung2013, Ville2018, Abuzarli2022, Steinhauer2022}. At longer times, deviations from integrability show up via quasi-particles interactions, leading to thermalization \cite{Regemortel2018, Duval2023}. Remarkably though, two main stages are thought to control the thermalization process itself. A fast, exponential decay of quasi-particles first ensures the onset of a local equilibrium, i.e., reached at short spatial scales. Later, a much slower relaxation is expected to take over, signaling the establishment of a \emph{global} equilibrium \cite{Lux2014}. This intriguing stage, still poorly understood, is governed by the few conservation laws and is often connected to diffusion \cite{Mukerjee2006, Lux2014, Friedman2020}, even though anomalous algebraic transport has been recently reported in several models \cite{Beijeren2012, Kulkarni2013, Kulkarni2015, Buchhold2015, Cao2018, Nardis2019, Bulchandani2020}. Global equilibration is generally difficult to characterize, both analytically and numerically, due to the particularly long time scales it involves. This is one of the reasons why most existing studies concern one-dimensional (1D) many-body systems.

\begin{figure}
\includegraphics[scale=0.53]{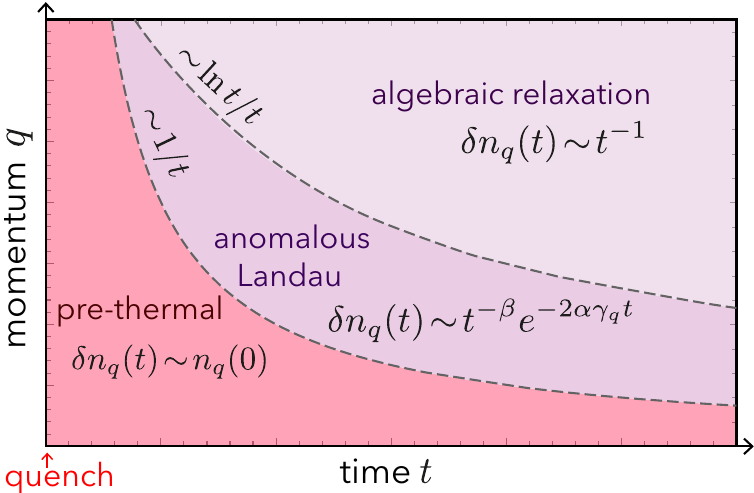}
\caption{\label{Fig:dynamics_phases}
Schematic momentum-time diagram of thermalization of excitations in a 2D Bose superfluid initially far from equilibrium.
 At short times, the dynamics is pre-thermal for all quasi-particle momenta $q$, i.e., the quasi-particle distribution $n_q(t)\!\simeq\!n_q(0)$ matches its post-quench value. 
As time progresses,  quasi-particles relax toward equilibrium, following a (Landau) exponential damping at a rate $\sim \gamma_q$ modified by nonlinear corrections $\propto t^{-\beta}$ with $\beta\simeq 1/2$.  Last, the fluid reaches its global equilibrium at the final temperature $T$ through an algebraic relaxation. This global equilibrium slowly establishes over space within a boundary $R(t)\!\sim\!t/\ln t$.
Here, $\delta n_q(t)\equiv n_q(t)\!-\!n_q^\text{th}$ is the deviation of $ n_q(t)$ from its thermal value $n_q^\text{th}$, and $\gamma_q$ is the quasi-particle scattering rate.}
\end{figure}
In this Letter, we undertake a precise spatio-temporal analysis of the long-time thermalization dynamics of a 2D, non-integrable many-body system:
a Bose superfluid undergoing a global interaction quench. While 2D superfluids  have  recently been the focus of out-of-equilibrium experiments involving ultra-cold atoms \cite{Hung2013, Ville2018, Sunami2022}  and optical fluids \cite{Abuzarli2022, Steinhauer2022}, their theoretical descriptions are still scarce. In superfluids, the dynamics following the pre-thermal stage is driven by the relaxation of low-lying phononic quasi-particles. 
Here, we characterize the spatio-temporal dynamics of this relaxation for a superfluid prepared in a far-from-equilibrium initial state, leveraging a quantum kinetic approach \cite{Landau_Lifshitz_Pitaevskii,KamenevBook,Furst2013}. This allows us to precisely identify the regions where local and global equilibria establish as time progresses, see Fig.~\ref{Fig:dynamics_phases}. Our study also reveals a striking difference with the dynamics of near-equilibrium initial states, for which the fast decay of quasi-particles is purely exponential, as conventionally described by linearized quantum Boltzmann equations \cite{Landau_Lifshitz_Pitaevskii, Micklitz2011, Lin2013}. Instead, we find that quenching from a far-from-equilibrium state introduces nonlinear algebraic contributions in the initial quasi-particle relaxation. Moreover, from high-precision numerical simulations of the kinetic equation up to long times, we manage to distinctly identify the onset of a slow, algebraic relaxation characteristic of global equilibration. This slow transport, which we find compatible with 2D diffusion, is intimately related to energy conservation and to the mechanism of Landau damping in 2D superfluids \cite{Pitaevskii1997, Giorgini1998, Chung2009}.

We consider a 2D interacting Bose gas with repulsive interactions, described by the many-body Hamiltonian $\hat{H} = \int d^2\br\big(\!- \frac{1}{2m}  \hat{\psi}^\dagger  \Delta_{\br} \hat{\psi} + \frac{g}{2} {\hat{\psi}}^{\dagger} {\hat{\psi}}^{\dagger}  \hat{\psi} \hat{\psi} \big)$, where $g>0$ is the interaction strength, and  the field operators $\smash{\hat{\psi}}$ satisfy the  bosonic canonical commutation rules. At low temperature and for weak interactions, the dynamics of the Bose gas is mainly governed by the phase fluctuations of the field. The latter is thus conveniently written as $\smash{\hat{\psi}(\br) =e^{i \hat{\theta}(\br)}\sqrt{ \hat{ \rho} (\br)} }$, where the density operator $\hat{\rho}(\br)=\rho_0+\delta\hat{\rho}(\br)$ displays small fluctuations $\delta\hat{\rho}\ll\rho_0$ around the mean gas density $\rho_0$~\cite{Popov1972, Mora2003}. Using this representation and expanding around $\delta\hat{\rho}$, we obtain a hydrodynamic formulation of the Hamiltonian, which we diagonalize in Fourier space by means of a Bogoliubov transformation: $\smash{{ \delta \hat{\rho}}_{\bq}^{\phantom{*}}\! =\! -\sqrt{{E_q}/{\epsilon_q}} ( \hat{a}_{\bq}^\dagger \! +\! \hat{a}_{-\bq}^{\phantom{\dagger}} )}$ and $\smash{\hat{\theta}_{\bq}^{\phantom{\dagger}}\!  =\! \frac{i}{2}\sqrt{{\epsilon_q}/{E_q}}  (  \hat{a}_{\bq}^\dagger  - \hat{a}_{-\bq}^{\phantom{\dagger}}  )}$,
where  the $\hat{a}_\bq$ are  annihilation operators of the superfluid's quasi-particles, $E_q\equiv \bq^2/(2m)$ and $\smash{\epsilon_q\equiv\sqrt{ E_q\left( E_q + 2 g \rho_0 \right)}}$ is the Bogoliubov dispersion. This leads to the effective, low-energy Hamiltonian
\begin{align}
\label{eq:H}
\hat{H} \!=\!\!  \int_\bq\,  \epsilon_q \,   \Big(\hat{a}_{\bq}^\dagger \hat{a}_{\bq}^{\phantom{\dagger}}\!+\!\frac{1}{2}\Big)
\!+\! \int_{\bp,\bq}\!\!\!  \Lambda_{p, q}^{\phantom{*}}  \left( \hat{a}_{\bp}^{\phantom{\dagger}} \hat{a}_{\bq}^{\phantom{\dagger}} \hat{a}_{\bp+\bq}^\dagger + \text{h.c.} \right).
\end{align}
where $\smash{\int_\bq\equiv \int d^2\bq/(2\pi)^2}$ and the function $\Lambda_{p,q}$ has been detailed in \cite{Buchhold2015, Buchhold2016, Duval2023}. The quadratic part of $\smash{\hat{H}}$ describes a gas of free quasi-particles with energy dispersion $\epsilon_q$, which are phonons at low energy: $\epsilon_q\simeq cq$ for $q\ll 1/\xi$, where $\smash{\xi\equiv \sqrt{1/4 g \rho_0 m}}$ is the healing length and $c = \sqrt{g \rho_0 /m}$ the speed of sound.  In a quench experiment, the quadratic term considered alone leads to a coherent dynamics of the superfluid excitations, $\hat{a}_\bq(t)=\hat{a}_\bq(0)\exp(-i\epsilon_q t)$, during which  the momentum distribution $n_q(t)\equiv\langle \hat{a}^\dagger_{\bq}(t)\hat{a}_{\bq}(t)\rangle=n_q(0)$ is time independent and matches the distribution $n_q(0)$ induced by the quench. Such dynamical regime is known as \emph{pre-thermalization}, and holds at short times only. The cubic term in Eq. (\ref{eq:H}) encodes interactions between quasi-particles, which are responsible for the relaxation of the superfluid toward equilibrium. In the following, we characterize this  relaxation through $n_q(t)$. For a weakly interacting superfluid, the latter is governed by a quantum kinetic equation derived from Eq. (\ref{eq:H}) \cite{Duval2023, Regemortel2018}:
\begin{align}\label{KE_born_nq}
	&\partial_{t} n_{q} =\!  \int_0^\infty \!dp \frac{3\sqrt{3}c}{4\pi\rho_0}p(p\!+\!q) \left[  n_{p+q} \left( n_p \!+\! n_q \!+\!1  \right)\! -\! n_p n_q  \right] \nonumber\\
	&+ \int_0^q \! dp \frac{3\sqrt{3}c}{8\pi\rho_0}p(p\!-\!q) [ n_p n_{q-p} \!-\! n_q \left( n_p \!+\! n_{q-p} \!+\!1 \right) ].
\end{align}
This kinetic equation has a single integral of motion, the total energy $\smash{\int_\bq cq\, n_q}$. The two collision integrals in the r.h.s of Eq. (\ref{KE_born_nq}) correspond to Beliaev and Landau scattering processes, which respectively describe the splitting $q\to(p,q-p)$ of the probe phonon of momentum $q$ into two phonons of momenta $p$ and $q-p$, and the recombination $(q,p)\to p+q$ of the probe phonon with another one. At long time,  $n_{q}(t)$ relaxes to the stationary solution $n_{q}^\text{th}=[\exp(\epsilon_q/T)-1]^{-1}$ of Eq. (\ref{KE_born_nq}), which is a
thermal Bose-Einstein distribution at a temperature $T$, determined from energy conservation.

\begin{figure}
\includegraphics[scale=0.6]{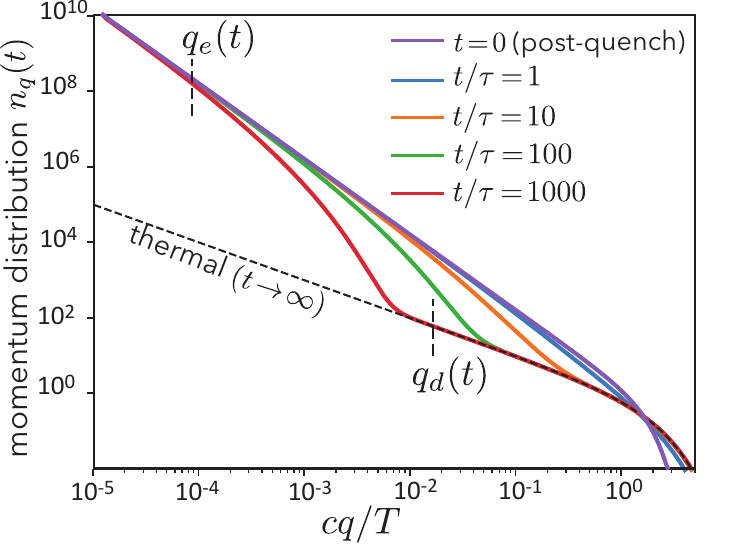}
\caption{\label{fig:Distributions} 
Quasi-particle distributions $n_q(t)$ vs. $q$ at increasing times, following an interaction quench in a 2D superfluid. 
The uppest purple curve is the post-quench distribution $n_q(0)$ that sets the time $t=0$. The lowest dashed curve is the thermal distribution $n_{q}^\text{th}=[\exp(cq/T)-1]^{-1}$ expected at long time. As time progresses, the dynamics reveals the emergence of three distinct regions separated by two characteristic momenta $q_e(t)$ and $q_d(t)$ that decay in time. Modes $q\!<\!q_e(t)$ have not yet relaxed, while modes $q\!>\!q_d(t)$ are almost thermalized. The relaxation at intermediate momenta $q_e(t)\!<\!q\!<\!q_d(t)$ is described by an anomalous Landau damping. In the legend, time is in units of $\tau=4\rho_0c^2/(\sqrt{3}\pi T^3)$, the typical quasi-particle lifetime at the thermal momentum $T/c$. 
}
\end{figure}

To quantitatively explore the thermalization dynamics of the superfluid, we have performed extensive numerical simulations of Eq. (\ref{KE_born_nq}) starting from a far-from-equilibrium superfluid state. This state was produced by tayloring an interaction quench $g_0\to g$ in a superfluid at an initial temperature $T_0$ \cite{Martone2018}, with $g_0,g,T_0$ chosen such that $T_0/g_0\rho_0\gg1$ and $T/g\rho_0\ll1$. In this way, the effective dispersion of the post-quench state is quadratic in a large region of the spectrum, i.e., $n_q(0)\simeq [\exp(q^2/2mT_0)-1]^{-1}$, in contrast to the dispersion of the final equilibrium state, which is essentially linear, i.e.,  $n_{q}^\text{th}=[\exp(cq/T)-1]^{-1}$ (see \cite{Supplemental} for details about the implementation of the post-quench state).  This defines a far-from-equilibrium post-quench state $n_q(0)$, which we use as an initial condition for Eq. (\ref{KE_born_nq}). 
The numerically computed distributions $n_q(t)$ are shown in Fig. \ref{fig:Distributions} at different times.
They reveal the existence of three distinct  dynamical regions separated by two characteristic momenta $q_e(t)$ and $q_d(t)$. At low momenta $q\!<\!q_e(t)$, first,  $n_q(t)\simeq n_q(0)$. 
This corresponds to the pre-thermal regime where  the phonon relaxation is largely ineffective. At intermediate momenta $q_e(t)\!<\!q\!<\!q_d(t)$, $n_q(t)$ deviates from $n_q(0)$ and rapidly evolves toward the final thermal state. At large momenta $q\!>\!q_d(t)$, finally, $n_q(t)$ seems to closely match $n_q^\text{th}$, signaling the establishment of a global equilibrium. In the following, we examine these two relaxation processes in more detail.
 \begin{figure}
\includegraphics[scale=0.82]{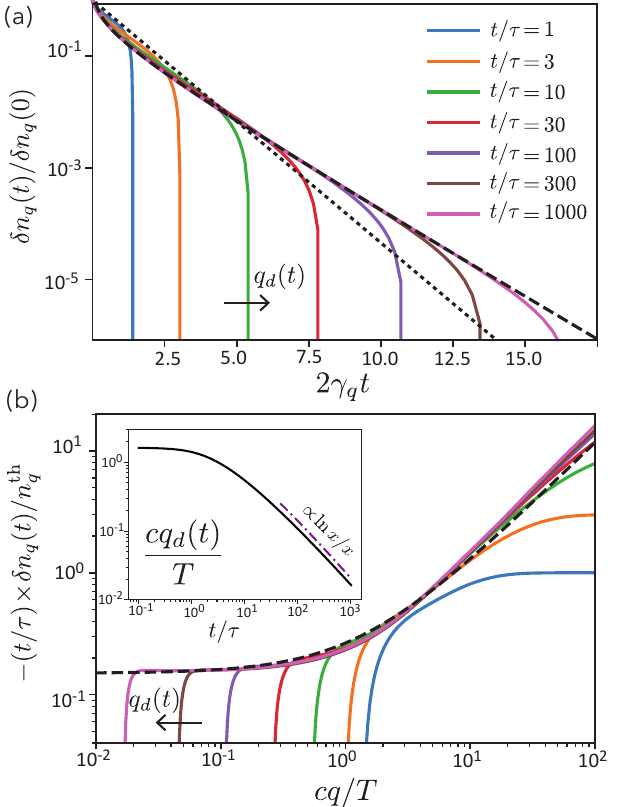}
\caption{\label{fig:rescaling}
(a) Deviation $\delta n_q(t)\equiv n_q(t)-n_q^\text{th}$ of the quasi-particle distribution from its equilibrium value as a function of the rescaled momentum $2\gamma_q t$, at different increasing times. Below $q_d(t)$, all distributions collapse on a single function of $\gamma_q t$. The latter is well approximated by Eq. (\ref{eq:IRansatz}) with $\beta\simeq 0.4$ and $\alpha\simeq 0.68$ as fit parameters (dashed curve).  The drop of the distributions at $q=q_e(t)$ stems from a change of sign of $\delta n_q(t)$. The dotted curve shows the purely exponential solution $\delta n_q(0)\exp(-2\gamma_q t)$ obtained from a simple linearization of the kinetic equation. 
(b) Rescaled distributions $t\times \delta n_q(t)$ vs. $q$ at increasing times, for $q>q_d(t)$. At long time all distributions collapse on a single curve, indicating  an algebraic scaling $\delta n_q(t)\propto 1/t$. The dashed curve is a plot of Eq. (\ref{eq:UVansatz}) without any adjustable parameter. The inset shows a numerical computation of $q_d(t)$ (solid curve), obtained by identifying $q_d(t)$ with the point where $\delta n_q(t)$ changes sign. At long time, $q_d(t)$ is well described by the asymptotic law (\ref{eq:qdt}) (dotted-dashed curve).
}
\end{figure}

\textit{Intermediate momenta --} To unveil the superfluid's dynamics at intermediate momenta, we plot in log-scale in Fig. \ref{fig:rescaling}(a) the  deviation $\delta n_q(t)\equiv n_q(t)-n_q^\text{th}$ of the distribution from equilibrium as a function of the rescaled momentum $2\gamma_q\times t$, for various times. Here $\smash{\gamma_q={\sqrt{3}\pi}q T^2/(8\rho_0 c)}$ is the Landau phonon relaxation rate, whose inverse can be seen as the lifetime of a quasiparticle of momentum $q$ \cite{Pitaevskii1997, Chung2009}. At small $q$, we find that all distributions in Fig. \ref{fig:rescaling}(a)  collapse on the same single curve, suggesting that $\delta n_q(t)$ is a function of $\gamma_q t$ only. Note that this collapse works only  below at a certain momentum where $\delta n_q(t)$ changes sign, a point that we identify with $q_d(t)$.
For $q\!<\! q_d(t)$, it turns out that a naive description using the linearized solution of (\ref{KE_born_nq}),  i.e. $n_q(t)\simeq \delta n_q(0)\exp(-2\gamma_q t)$,  fails to capture the numerical results [see dotted curve in Fig. \ref{fig:rescaling}(a)]. The reason is that for a far-from-equilibrium state, low-momentum modes are strongly populated, so the nonlinear contributions in Eq. (\ref{KE_born_nq}) can never be neglected. From a careful analysis of the kinetic equation detailed in \cite{Supplemental}, we find that a distribution of the form
\begin{equation}
\label{eq:IRansatz}
\delta n_q(t)\sim \delta n_q(0) \frac{\exp(-2\alpha \gamma_q t)}{(2\gamma_q t)^\beta},\ \ \ \ q_e(t)\!<\!q\!<\!q_d(t),
\end{equation} 
with $\beta=1/2$ and $\alpha=0.72$, is an approximate solution of Eq. (\ref{KE_born_nq}) at intermediate momenta and long times.
We refer to Eq. (\ref{eq:IRansatz}) as an \emph{anomalous} Landau damping of quasi-particle modes.
This relaxation is confirmed by a direct fit of the numerical data of Fig. \ref{fig:rescaling}(a) to Eq. (\ref{eq:IRansatz}), which  yields values $\beta=0.4$ and $\alpha=0.68$ close to the theoretical estimates. 
Equation (\ref{eq:IRansatz}) also allows us to identify the momentum scale $q_e(t)$ beyond which the system escapes the pre-thermal regime. Indeed, by definition   $q_e(t)$ obeys $\delta n_{q_e}(t)\sim \delta n_{q_e}(0)$, which leads to:
\begin{equation}
\label{eq:qet}
q_e(t)\sim \frac{T}{c}\frac{\tau}{t},
\end{equation}
where we have introduced $\tau=4\rho_0c^2/(\sqrt{3}\pi T^3)$, the quasi-particle lifetime at the thermal momentum $T/c$. Physically, 
within the intermediate momentum range where Eq. (\ref{eq:IRansatz}) holds, we expect the system to get close to equilibrium,  though with fluctuations  differing from those of a genuine thermal state \cite{Lux2014}. This implies, in particular, that at scales $q_d(t)^{-1}<R<q_e(t)^{-1}$ the system can at best be described by a local equilibrium state.

\textit{Large momenta --} We now turn our attention to the large-momentum regime $q>q_d(t)$ in Fig. \ref{fig:Distributions}, where $n_q(t)$  seemingly becomes very close to its thermal value $n_q^{\text{th}}$. We attribute the spatio-temporal crossover corresponding to this change of behavior to the establishment of a global equilibrium, wherein the system's fluctuations should  resemble those of a genuine superfluid thermal state at the temperature $T$. To gain more insight on the dynamics at large momenta, we show in Fig. \ref{fig:rescaling}(b) the rescaled distribution $-t/\tau\times\delta n_q(t)/n_q^\text{th}$ as a function of $q$. We observe that for $q>q_d(t)$, these rescaled distributions collapse on a single curve at long time, suggesting a quasi-particle distribution of the form $\delta n_q(t)/n_q^{\text{th}}\propto \tau/t$. 

To clarify the origin of this algebraic tail, it is instructive to look at the constraint of energy conservation, $\int_\bq\, cq\, \delta n_q(t)=0$, which holds at all times.  By segmenting the integral over $q$ across the three dynamical regions identified earlier, we obtain a relation for the relative energy of the ultraviolet (UV) modes $q>q_d(t)$:
\begin{equation}
\label{eq:energyconservation}
\delta E_\text{UV}(t)\!\equiv\!\int_{q_d(t)}^\infty \!\!\!dq\,cq^2 \delta n_q(t)\!=\!-\delta E_\text{P}(t)\!-\!\delta E_\text{L}(t).
\end{equation}
The right-hand side contains the sum of the relative energy $\delta E_\text{P}$ of pre-thermal modes, and the relative energy $\delta E_\text{L}$ of modes experiencing anomalous Landau damping.  By definition,
$\delta E_\text{P}(t)\equiv \int_0^{q_e} dq\, cq^2 \delta n_q(0)\sim 2mT_0 q_e(t)\sim \tau/t$, where Eq. (\ref{eq:qet}) has been used.
On the other hand,  the relative energy of modes relaxing via Landau damping is  $\delta E_\text{L}(t)\equiv \int_{q_e}^{q_d} dq\, cq^2 \delta n_q(t)$, where at long time $\delta n_q(t)$ follows Eq. (\ref{eq:IRansatz}). This leads to $\delta E_\text{P}(t)\simeq \int_{q_e}^\infty dq\,cq^2\delta n_q(0)\exp(-2\alpha \gamma_q t)/(2\gamma_q t)^\beta\sim \tau/t$.  Inserting these two results into  Eq. (\ref{eq:energyconservation}) and assuming a separable form $\delta n_q(t)=f(t/\tau)g(q)$ for the quasi-particle distribution in the UV region, we immediately obtain $f(t/\tau)\sim \tau/t$, thus confirming the numerical observation of Fig. \ref{fig:rescaling}(b). A further quantitative analysis of the kinetic equation detailed in \cite{Supplemental} allows us to also compute $g(q)$, yielding:
\begin{equation}
\label{eq:UVansatz}
\frac{\delta n_q(t)}{n_q^\text{th}}\sim \frac{\tau}{t}\Big[1\!+\!\frac{cq/T}{4+\pi^4/30\zeta(3)}+\mathcal{O}(q^2)\Big], \ \ q\!>\!q_d(t).
\end{equation} 
Plotting this prediction on top of the numerical distributions in Fig. \ref{fig:rescaling}(b), we find an excellent agreement. Moreover, by extrapolating Eqs. (\ref{eq:IRansatz}) and (\ref{eq:UVansatz}) at $q=q_d(t)$, we get an estimation for the characteristic momentum scale $q_d(t)$ separating the regimes of  anomalous Landau damping and algebraic relaxation at times $t\gg\tau$:
\begin{equation}
\label{eq:qdt}
q_d(t)\sim \frac{T}{c}\frac{\ln(t/\tau)}{t/\tau}.
\end{equation}
In the inset of Fig. \ref{fig:rescaling}(b), we show a numerical computation of $q_d(t)$, obtained by identifying $q_d(t)$ with the point where $n_q(t)-n_q^\text{th}$ changes sign. At long time, $q_d(t)$ is well described by the asymptotic law (\ref{eq:qdt}). Notice that the temporal decay of $q_d(t)$ is relatively slow, which indicates that reaching a global equilibrium across large spatial regions is a lengthy process. The dynamical relations (\ref{eq:IRansatz}), (\ref{eq:qet}), (\ref{eq:UVansatz}) and (\ref{eq:qdt}) are summarized by the momentum-time phase diagram  in Fig. \ref{Fig:dynamics_phases}.

At this stage, a few comments are in order. First,
the long tail $\propto1/t=1/t^{d/2}$ in Eq. (\ref{eq:UVansatz}) is imposed by  energy conservation. It is compatible with a mechanism of 2D diffusive transport of the superfluid's excitations at long time \cite{Friedman2020}, although further analysis would be needed to confirm this connection. Second, we note that this algebraic tail is particularly pronounced in our system, which appears to stem from the far-from-equilibrium nature of the initial state. Indeed, when conducting similar simulations with a near-equilibrium initial distribution $n_q(0)\sim T_0/(cq)$, we could also identify a slower relaxation at long time, but without clear algebraic scaling, akin to the results reported in \cite{Lux2014} for a 1D Bose Hubbard model. 
Third, it is interesting to note that the $1/t$ tail also crucially depends on the  linear dependence on momentum of the Landau damping rate, $\gamma_q\propto q$. This condition, for example, was  shown to break in the 1D non-equilibrium L\"uttinger liquid \cite{Beijeren2012, Kulkarni2013, Kulkarni2015} where, unlike in two dimensions,  phonon scattering processes are resonant, i.e. they simultaneously conserve energy and momentum. This yields a different scaling $\gamma_q\propto q^{3/2}$ \cite{Andreev1980, Grassberger2002, Beijeren2012, Kulkarni2013, Kulkarni2015, Buchhold2015}, from which 
we infer $\delta E_P(t)\sim 1/t^{2/3}$. Therefore, according to the conservation law (\ref{eq:energyconservation}), in 1D the distribution scales as $\delta n_q(t)\sim 1/t^{2/3}$  instead of $1/t$ at long time, implying a Kardar-Parisi-Zhang rather than diffusive-like dynamics. 
\begin{figure}
\includegraphics[scale=1.]{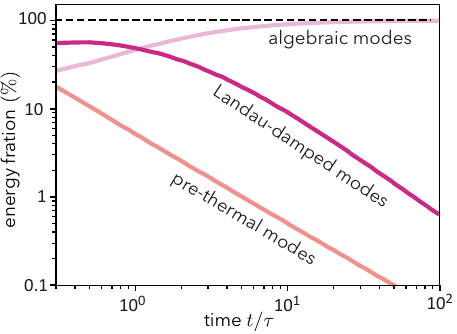}
\caption{\label{fig:energyfraction}
Total energy fractions  of pre-thermal, Landau-damped and algebraically-relaxing modes as a function of time. The sum of the three fractions equals 1 due to energy conservation. At  times $t/\tau\gg 1$, both fractions of pre-thermal and Landau-damped modes decay as $\tau/t$.}
\end{figure}

While the distributions in Fig. \ref{fig:Distributions} provide a spatially-resolved analysis of the thermalization process, it is also instructive to examine the \emph{total} energy fractions of pre-thermal, Landau-damped and algebraically-relaxing modes. These fractions are  defined as $(\int_I dq c q^2n_q)/(\int_0^\infty dq c q^2n_q)$, where the interval of integration is $I=(0,q_e)$ for pre-thermal modes, $I=(q_e,q_d)$ for Landau-damped modes and $I=(q_d,\infty)$ for algebraic modes. These fractions are displayed in Fig. \ref{fig:energyfraction} as a function of time. As expected, the fraction of pre-thermal modes decays fast beyond the quasi-particle lifetime $\tau$. The fraction of Landau-damped modes, on the other hand, requires $\sim 100\tau$ to become smaller than $1\%$, indicating that $99\%$ of the modes are thermalized at this time scale.

In this work, we have provided a first, in-depth exploration of the spatio-temporal dynamics of thermalization in 2D Bose superfluids. Starting from a far-from-equilibrium state, we have identified two distinct relaxation regimes: an initial anomalous damping of excitations, followed by a slow dynamics with a pronounced algebraic tail. An interesting extension of this work would be to explore the long-time thermalization of Bose gases quenched across their superfluid transition, as realized in recent experiments \cite{Abuzarli2022, Sunami2022, Glidden2021}. 

This work has benefited from the financial support of Agence Nationale de la Recherche (ANR), France, under Grants  Nos.~ANR-19-CE30-0028-01 CONFOCAL for NC, and ANR-18-CE30-0017 MANYLOK for CD.

\end{document}